\documentclass[%
 reprint,
 amsmath,amssymb,
 aps,
]{revtex4-2}
\usepackage{comment}
\usepackage{graphicx}% Include figure files
\usepackage{dcolumn}% Align table columns on decimal point
\usepackage{bm}% bold math
\usepackage[normalem]{ulem}
\usepackage{xcolor}
\usepackage{siunitx}

\DeclareRobustCommand{\Erase}{\bgroup\markoverwith{\textcolor{red}{\rule[.5ex]{2pt}{0.4pt}}}\ULon}

\begin{document}

%\preprint{APS/123-QED}

\title{Sliding of cylindrical shell into a rigid hole}% Force line breaks with \\

\author{Yukiho Matsumoto$^1$}
\author{Keisuke Yoshida$^2$}
\author{Tomohiko G. Sano$^{1,3}$}%
\email{sano@mech.keio.ac.jp}

\affiliation{$^1$School of Integrated Design Engineering, Graduate School of Science and Technology, Keio University, 3-14-1 Hiyoshi, Yokohama, Kanagawa, 2238522, Japan.}
\affiliation{$^2$Research Organization of Science and Technology, Ritsumeikan University, Kusatsu, Shiga 525-8577, Japan}
\affiliation{$^3$Department of Mechanical Engineering, Faculty of Science and Technology, Keio University, 3-14-1 Hiyoshi, Yokohama, Kanagawa, 2238522, Japan.
}%

\date{\today}% It is always \today, today,

\begin{abstract}

Fitting two different materials is a versatile methodology in manufacturing complex structures. One of the canonical models for fitting is the snap-fit model, in which flexible materials and rigid structures are assembled by pushing their interlocking components together. 
The assembly via snap-fit is often accompanied by large deformations of flexible structures and abrupt force drops, highlighting the role of elasticity, geometry, and contact friction. 
Despite several model studies revealing fundamental mechanics for snap-fit, the current snap-fit design relies on prototyping and empirical rules. In this paper, we analyze a snap-fit model in which a naturally curved beam slips into a rigid hole. We construct an analytical model based on the theory of elastica with contact friction and demonstrate that its predictions are in excellent quantitative agreement with both simulations and experiments. We find three distinct sliding modes: Folding, Pinning, and Unfolding. The classification is systematically organized in a phase diagram based on the geometric parameters of the shells and the hole. Our study complements existing approaches by providing a predictive framework for contact-based structures that involve friction, elasticity, and geometry, and sheds light on a unified understanding of the interactions between an elastic and a rigid body.

\end{abstract}

\maketitle

\section{Introduction}

The assembly and disassembly of two structures are essential for the emergence of novel functionality in both natural and artificial systems. These operations are observed across a wide range of length scales, including ligand-receptor interactions in biochemistry~\cite{moy1994intermolecular}, the assembly of industrial products~\cite{benson1982stick}, docking between spacecrafts in free flight~\cite{romano2007laboratory}, and even children's toys~\cite{yoshida2020mechanics}. Fitting two different structures relies on mechanical and geometric nonlinearities as well as unique interactions at each length scale. 

One of the canonical models for fitting structures is snap-fit, in which one component is inserted into the other, producing a snapping sound. Many snap-fit structures require a greater force for detachment than for fitting. As such, snap-fitted structures are maintained in a configuration different from their natural state by external forces and constraints. The necessary force for fitting/detaching them is determined by a subtle balance between elasticity, geometry, and friction~\cite{benson1982stick,yoshida2020mechanics}. Despite the widespread practical use of snap-fits, their design in product development often relies on empirical data and experiential knowledge. Recently, Yoshida \& Wada performed a model system study for snap-fit, where the set of an open cylindrical shell and a rigid cylinder acts as a snap-fit, by combining experiments, simulations, and theory~\cite{yoshida2020mechanics}. When a cylindrical shell is compressed against a cylinder, the shell either abruptly or smoothly embraces the cylinder, depending on its geometry and contact conditions.

Understanding the mechanics of snap-fit involves the interplay of friction, elasticity, and geometry~\cite{Landau1980}. Even if the structure is simple, predictions of its deformation in the presence of contact require extensive efforts that combine experiments and theory~\cite{sano2017slip, Grandgeorge2021cap}. The simple model system studies, which are minimal models simplifying the complexities in real engineering problems~\cite{Powers2010, Audoly2010}, help predict the mechanics of knots, yarns, and knits~\cite{ivey1999knot,naik2000twisted,audoly2007elastic, Crassous2024}, the performance of mechanical metamaterials~\cite{Rafsanjani2018}, energy absorption devices~\cite{sano2023randomly}, and soft actuators~\cite{gorissen2020inflatable, Yang2021, abe2024snap}. The contact condition of structures is crucial in predicting both their statics and dynamics.

To demonstrate that friction is essential for slender structures in contact, we introduce previous research~\cite{yoshida2024indentation} that is also relevant to our study. Yoshida \& Wada systematically perform mechanical testing of open cylindrical shells against a flat substrate~\cite{yoshida2024indentation}. The deformation of the cylindrical shell is classified into three types, \textit{Folding, Unfolding, and Pinning}, depending on the geometry of the shell and the friction coefficient between the shell and the substrate. When the friction coefficient is small, the shell closes inward (Folding) or opens outward (Unfolding), depending on its opening angle. On the other hand, when the friction is sufficiently large, the tips of the shell are fixed and ``pinned," which is classified as Pinning. It should be stressed that both geometry and friction contribute to the classification of deformation modes of the shell.

Here, we study the mechanics of an open cylindrical shell when it fits and slides into a rigid hole of open-circle shape (Fig.~\ref{fig:1}(a)). By indenting the open cylindrical shell to `drip' it into the hole, we observe three distinct sliding behaviors of the shell, which are classified as \textit{Folding, Pinning, and Unfolding}, inspired by Yoshida \& Wada~\cite{yoshida2024indentation}. We systematically conduct experiments and simulations to understand the mechanisms underlying these sliding behaviors, supplemented by the theory of elastica~\cite{Wang1986, He1997, Nizette1999, Plaut2011} that incorporates contact friction. The analytical predictions are in excellent agreement with experiments and simulations and uncover the roles of friction coefficients in sliding behavior. 
Our model system is complementary to the self-encapsulation of elastic beams, analogous to dripping liquid~\cite{Bosi2015}. 
Our findings would push forward the understanding of the mechanics of a cylindrical shell~\cite{yoshida2024indentation} toward the case when pressed by a complex-shaped indenter, as well as when cylindrical shells are in contact with random orientations relevant in stacked configurations~\cite{sano2023randomly}.

The remainder of this paper is organized as follows. In Sec.~\ref{sec:Problem definition}, we provide the problem setting addressed in this study, including experimental and simulation methods. In Sec.~\ref{sec:Sliding modes of elastic shells}, we analyze the mechanisms of three distinct sliding modes during the process of an open cylindrical shell fitting into a hole, and summarize experimental and numerical results as a phase diagram. In Sec.~\ref{sec:Elastica model}, we introduce the elastica model as a theoretical framework to classify the sliding behaviors presented in Sec.~\ref{sec:Sliding modes of elastic shells}. In Sec.~\ref{sec:Effects of friction}, we clarify the role of friction coefficients in sliding behaviors (both hole and indenter) based on the elastica model. In Sec .~\ref {sec:Discussion and Summary}, we discuss the results presented in this paper.

\section{Problem definition}\label{sec:Problem definition}
\begin{figure}
    \centering
    \includegraphics[width=\linewidth]{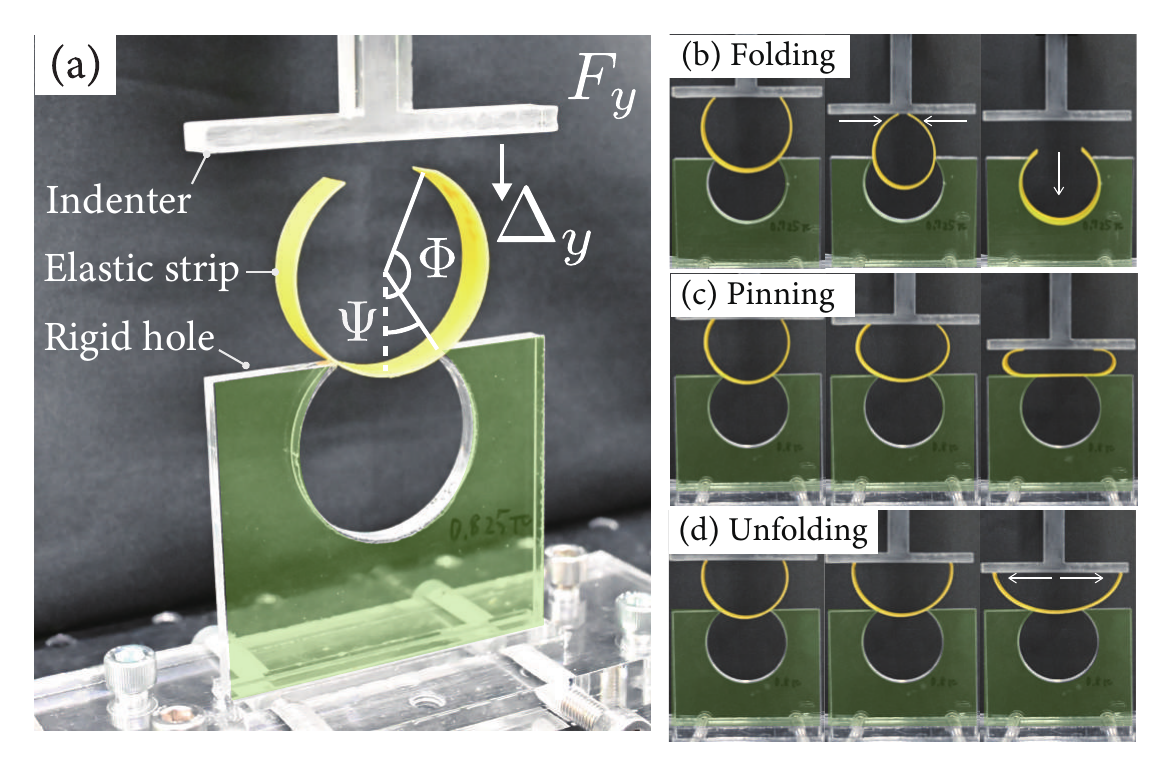}
    \caption{
        An elastic strip with spontaneous curvature is inserted into a hole in a rigid body. (a) Experimental setup. The naturally-curved elastic strip of total length, $2L$, and radius of curvature, $R$, is pressed by a rigid indenter from above by $\Delta_y$ and is slid into the rigid hole underneath. (b)–(d) Snapshots of three different sliding modes obtained from the experiments for three characteristic configurations: (b)~Folding, (c)~Pinning, and (d)~Unfolding
        }
    \label{fig:1}
\end{figure}

Our model system consists of a naturally-curved elastic strip of open cylindrical geometry (elastic shell), a flat rigid indenter, and a rigid substrate with a circular-segment-shaped notch, which we call a rigid hole (Fig.~\ref{fig:1}(a)). An elastic shell has a radius of curvature $R$, thickness $t$, width $b$, and total length $2L$. The bending performance of the shell is characterized by the bending modulus, $B = Et^3b/12$. The rigid substrate placed below the shell has a hole of a (significant) circular segment, whose radius and chord-length are respectively given by $R$ and $2W$.
The relevant dimensionless parameters in our model system are the (half of) opening angles of the shell, $\Phi$, and the hole, $\Psi$, which are calculated geometrically as $\Phi\equiv L/R$ and $\Psi\equiv \sin^{-1}(W/R)$ as schematically shown in Fig.~\ref{fig:1}(a). The elastic shell is carefully placed over the hole so that the symmetric axes of the shell and the hole are aligned, and the indenter is lowered by $\Delta_y$ at a sufficiently low speed to measure the force for the shell to slide into the hole, $F_y$.
We assume that the shell interacts with the indenter and the edge of the hole via contact friction satisfying the Amontons-Coulomb law~\cite{persson2013}, whose static friction coefficients are denoted as $\mu_{\mathrm i}$ and $\mu_{\mathrm e}$, respectively. 

\subsection{Experiments}

Elastic shells in our experiments are cast in a thermoplastic way from naturally straight ribbons made of glycol-modified polyethylene terephthalate, as Ref.~\cite{sano2023randomly}. The straight ribbon is laser-cut to a length of $2L$ and a width of $b = 9$~mm from a flat sheet ($ E = 2.2$~GPa, $t = 0.5$~mm, Shim Stock, Artus Corporation, USA). The total length varies as $2L\in[124,198]$~mm, where the opening angle, $\Phi$, spans $0.55\pi\leq\Phi\leq0.90\pi$. The mold for the open shells consists of two. The acrylic outer mold has a circular hole of radius $R$, while the inner circular mold, with identical geometry to the hole, is made of silicone elastomer (Smooth-on, USA). The naturally straight ribbon is sandwiched between the outer and inner molds and is bent to form an arc of radius $R$. The set of a ribbon and molds is placed into the hot water (95$^{\circ}$C) for more than 15 minutes and is cooled in cold water (20$^{\circ}$C) and then demolded. The natural radius of the curvature of the shell $R$ is fixed throughout.

The rigid substrate with a notch and the indenter are laser-cut from acrylic plates of thicknesses $5$~mm and $10$~mm, respectively. The geometry of the hole is characterized by the radius and opening angle, $(R, \Psi)$. The radius of the hole is chosen to be identical to the radius of natural curvature of the shell, $R$ (fixed throughout as $19$~mm), while the opening angle is systematically varied as $0.1\pi\leq\Psi\leq0.35\pi$. 
Even when the radius of the hole is different from $R$, the onset of the sliding behavior is not affected as long as the width of the hole, $W$, is identical.
The indenter is T-shaped, which is clamped to the load cell of the force-testing machine (EZ-LX, Shimadzu, Japan).

The procedure for the experiment is as follows. We set the elastic shell on the rigid substrate so that the outer surface of the shell touches the edge of the hole and its extremities point upward (Fig.~\ref{fig:1}(a)). The shell is placed symmetrically with respect to the loading direction as much as possible, to avoid the asymmetric sliding and rolling on the substrate. The indenter is slowly lowered from above at a constant speed $1~\mathrm{mm/s}$. The indenter displacement and reaction force are simultaneously recorded to evaluate the force-displacement relationship of the shell during mechanical testing. Two acrylic plates of 1~mm are placed in front and behind the shell and substrate, separated by 11~mm, to prevent out-of-plane deformation.

\subsection{Numerical Simulation}

To understand the mechanism of the sliding phenomena, we perform the centerline-based simulations assuming that the deformation of the shell is planar and can be regarded as a naturally-curved beam~\cite{chirico1994kinetics, sano2017slip, yoshida2020mechanics}. We use the Cartesian $x$-$y$ coordinate system so that the original point and the $y$-axis correspond to the center of the arc in the natural configuration and the symmetric axis of the shell, respectively. We express the deformation of the shell by the centerline position of the shell, $\bm{r}(s) = (x(s), y(s))$, assuming that the shell is sufficiently thin as $t/L\ll1$, where $s$ represents the arclength of the shell centerline measured from the intersection of symmetric axis ($x = 0$) and centerline in the reference configuration ($-L\leq s\leq L$). Within the discrete simulation framework, the centerline position of the beam, $\bm{r}(s)$, is discretized into a set of connected beads via a stiff linear spring as $\bm{r}_i (i = 1,2,\cdots, N)$. 

The equilibrium configuration of the beam centerline is determined by minimizing the elastic bending energy, expressed as a function of the position of the beads, subject to boundary conditions. We employ the elastic bending energy density of the beam, $\epsilon_b$, as a quadratic function of curvature difference from the natural configuration as, $\epsilon_b = (B/2)(\kappa(s) - R^{-1})^2$, where the curvature of the beam is calculated as $\kappa(s) = d\theta/ds$ with the tangent angle of the centerline, $\theta(s)$, obtained as $\hat{\bm{t}}(s) = \bm{r}'(s) = (\cos\theta(s), \sin\theta(s))$. The total elastic energy, $E_b\equiv\int_{-L}^{L} \epsilon_bds$, is discretized and is rewritten as a function of the set of $\{\bm{r}_i\}_{i = 1,\cdots, N}$. The bending force acting on the $i$-th node is then calculated by differentiating the energy as $-\partial E_b/\partial \bm{r}_i$. To realize the inextensibility of the centerline approximately, the adjacent beads are connected via a stiff linear spring. The positions of the beads are updated according to their (overdamped) equations of motion. This general framework is used to simulate filament-like structures such as rods and ribbons~\cite{bergou2008discrete, yoshida2020mechanics, yoshida2024indentation, sano2022kirchhoff}, and the simulated large deformations are consistent with experiments. 

The normal and tangential contact interactions between the indenter and the substrate are modeled by calculating the overlaps between the beads and the indenter. Let $H$ be the height of the indenter. Upon compression, the height, $H$, decreases as $H = H_0 - \Delta_y$ with the initial height $H_0$ defined as the height where the heights of the indentor and beads at the tips are identical ($F_{y} = 0$). As the indenter is lowered, the vertical overlaps are calculated for all bead locations. When they overlap as $\delta_i \equiv y_i - H > 0$, the repulsive force, $N_i$, linear to $\delta_i$, is applied to the bead. The proportionality factor is chosen to be sufficiently large so that the penetration of the indenter against the shell is sufficiently small. The tangential force is modeled by the Amontons-Coulomb law. The bead in contact is immobile if the magnitude of the tangential force calculated from the shell configuration, $T_i$, is lower than the maximum static friction force, $\mu_iN_i$, while the bead slips with the kinetic friction force $\pm\mu_i'N_i$, where $\mu_i$ and $\mu_i' (\equiv 0.8\mu_i)$ represent the static and kinetic friction coefficients, respectively~\cite{sano2017slip}. The reaction force, $F_{y}$ (measured in experiments), is calculated from the sum of the repulsive forces, $N_i$, acting on the indenter via the action-reaction law. 

When the shell comes into contact with the hole, the edges of the hole exert point forces at contact positions (See Fig.~\ref{fig:1}(a)). They still interact by points even if the loading test progresses, as long as the shell snaps into the hole completely or is squashed (detailed later). Hence, for simplicity, we consider only interactions at the edges of the holes, not the entire surface, and compute their distances to determine the contact forces. Before loading, the contact nodes are set to share the same coordinates as the edges, with contact forces set to zero. The initial contact nodes are connected to the corresponding edges via a linear stiff spring (whose spring constants are set to be large such that their penetrations are negligible). The normal, $F_n$, and tangential components (friction), $F_t$, of the contact forces are calculated by decomposing the forces via springs along the normals and tangents of the deformed configuration of the shell, respectively. Once the magnitude of the friction force, $|F_t|$, exceeds the maximum, $\mu_{\mathrm e} F_n$, the node slips, and we terminate the simulation. To further simulate the post-slip behavior on the edge, one would need to improve the contact modeling for the precise geometry of the hole shape \textit{e.g.,} by applying an appropriate dry friction model~\cite{romero2021physical, sano2023randomly}. However, our current modeling still yields good agreement with experiments for the force-displacement curve and the classification of sliding modes, as detailed later.

\section{Sliding modes of elastic shells}\label{sec:Sliding modes of elastic shells}

\begin{figure}[!h]
    \centering
    \includegraphics[width=\linewidth]{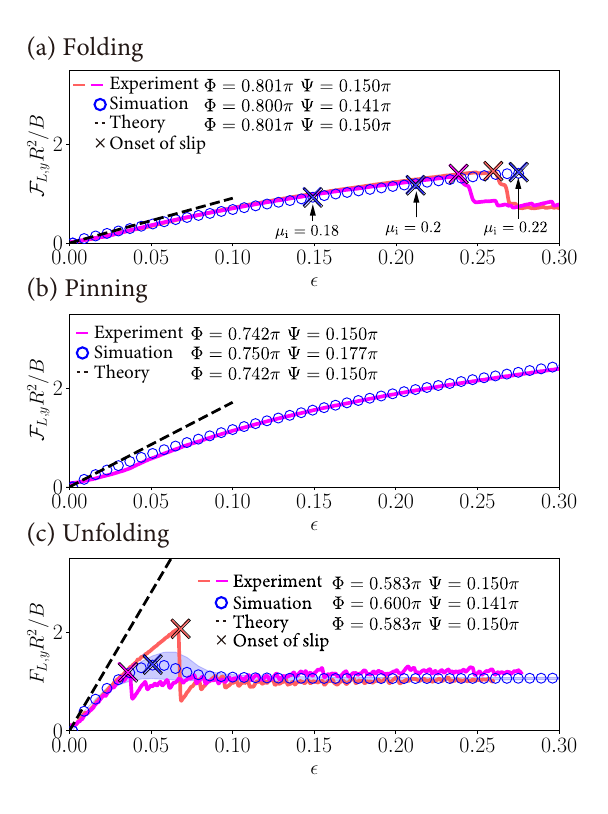}
    \caption{
         Rescaled force-displacement curve measured in experiments(solid line) and simulations($\bigcirc$). The dashed lines represent the linear response theory. (a) Folding, $(\Phi, \Psi) \approx (0.80\pi, 0.15\pi)$, (b) Pinning, $(\Phi, \Psi) \approx (0.6\pi, 0.15\pi)$, and (c) Unfolding, $(\Phi, \Psi) \approx (0.75\pi, 0.15\pi)$. In both experiments and simulations ($\mu_{\mathrm i}=\mu_{\mathrm e}=0.2$), cross symbols ($\times$) indicate the slip onset at the tip, $s = L$, corresponding to the condition such that the friction force at the indenter reaches the maximum as $|\mathcal{F}_{L,x}/\mathcal{F}_{L, y}| = \mu_{\mathrm i}$. 
         The simulation results for different friction coefficients of $\mu_{\rm i}~(0.18\leq\mu_{\rm i}\leq0.22)$ are shown in (a) and (c). (a)~The $\times$ marks represent the corresponding slip onset, while (c)~the corresponding force displacement curves are shown as the upper and lower bounds of the shaded region. 
         The dashed lines represent the prediction from the linear response theory. 
         }
     \label{fig:2}
\end{figure}

The deformation of the shell depends on the dimensionless geometric parameters; the opening angles of the shell $\Phi$ and the hole $\Psi$. We find three distinct sliding modes: Folding, Pinning, and Unfolding. The shell slips either on the indenter or on the edge of the hole, followed by opening or closing deformation, as detailed below. We classify the deformation mode based on the rescaled force-displacement curve, where the indenting force, $F_y$, and the displacement, $\Delta_y$, are rescaled by their characteristic quantities as $F_y R^2/B$($=2\mathcal{F}_{L,y}R^2/B$, the normal contact force for each tip, $\mathcal{F}_{L,y}$, is introduced as $\mathcal{F}_{L,y}\equiv F_y/2$ for theoretical purpose. See Sec.~\ref{sec:Elastica model} for details.) 
and $\epsilon\equiv\Delta_y/R$, respectively (Fig.~\ref{fig:2}).

\begin{figure}
    \centering
    \includegraphics[width=\linewidth]{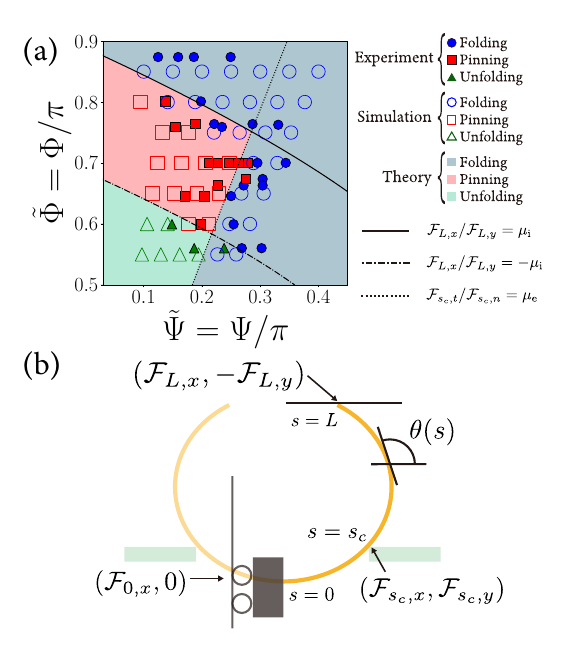}
    \caption{
        (a)Phase diagram constructed from theory, simulations, and experiment summarized on the $(\tilde\Psi,\tilde\Phi)$ parameter space. The background color and phase boundaries are predicted from linear-response theoretical analysis. Analytical phase boundaries are indicated by different line styles: solid and chain lines for slip in the closing and opening directions at the indenter as $\mathcal{F}_{L,x}/\mathcal{F}_{L,y} = \pm \mu_{\mathrm i}$, respectively, and dotted lines for slipping at the the edge of the hole as $\mathcal{F}_{s_c,t}/\mathcal{F}_{s_c,n} = \mu_{\mathrm e}$. Both theoretical and simulation results are obtained with the friction coefficients $\mu_{\mathrm i} = \mu_{\mathrm e} = 0.2$. The corresponding filled and empty symbols represent the experimental and simulation results, respectively. (b) Coordinate system and variable definitions for theoretical analysis. Due to the left-right symmetry of the natural shape, half of the elastic strip is analyzed. 
    }
    \label{fig:3}
\end{figure}

When the shell has a large opening angle $\Phi$ and the hole has a large opening angle $\Psi$, the shell folds (Figs.~1(b) and 2(a)). The shell folds inward, allowing it to eventually fit into the hole, which we call Folding mode. 
As the indenter is lowered, each tip exerts normal and tangential forces on the indenter and the contact points do against the edges, whose contact forces on the right half of the shell as $\bm{\mathcal{F}}_{L}=(\mathcal{F}_{L,x}, \mathcal{F}_{L,y})$ and $\bm{\mathcal{F}}_{s_c}=(\mathcal{F}_{s_c,x},\mathcal{F}_{s_c,y})$, respectively (See Fig.~\ref{fig:3}(b) for the schematic). We show the typical force-displacement curve for the folding mode in Fig.~\ref{fig:2}(a). The normal force at the indenter ($F_y (= 2\mathcal{F}_{L,y} $)) increases along with $\epsilon$ up to the slippage between the tip and the indenter (highlighted by $\times$ and force drop in Fig.~\ref{fig:2}(a)). As soon as the tip slips, the width of the shell decreases, and the contact angle at the edge is inclined to satisfy the slip condition at the edge, leading to the folding deformation as shown in the experimental curves for two independent trials of Fig.~\ref{fig:1}(b). We classify the deformation mode as Folding, if we observe the above deformation up to the prescribed maximum value of $\epsilon$ as $\epsilon = 0.3$.

In contrast, when both $\Phi$ and $\Psi$ are small, the shell undergoes Unfolding mode (Fig.~\ref{fig:1}(d) and~\ref{fig:2}(c)). The opposing $x$-direction force at the indenter exceeds the static friction. As the shell is compressed, it opens, increasing its width and preventing it from fitting into the hole. Once both tips of the shell slip completely (highlighted by $\times$ marks in Fig.~\ref{fig:2}(c)), the force remains almost constant. 

A distinct behavior appears for intermediate values of both $\Phi$ and $\Psi$, where neither folding nor unfolding dominates (Fig.~\ref{fig:1}(c) and~\ref{fig:2}(b)). The tips of the shell are pinned to the indenter, from which we call the Pinning mode. Unlike the previous two deformation modes, the friction forces never exceed the maximum static friction force. As a result, the force increases continuously without any abrupt drops (Fig.~\ref{fig:2}(b)). When the indentation is large enough, the tips of the shell even lift from the indenter, and the outer surface of the shell contacts with the indenter (See Fig.~\ref{fig:1}(c)).

We observe qualitatively consistent force displacement curves in our numerical simulations for a given set of geometric parameters $(\Phi, \Psi)$ as Fig.~\ref{fig:2}(a)-(c). In Fig.~\ref{fig:2}(a), we compare the experimental force curve for the Folding mode with the simulation results for the friction coefficients of the indenter as $\mu_{\rm i} = 0.18,0.20,0.22$. We find that the larger $\epsilon$ is required to fold the shell for larger $\mu_{\rm i}$. The comparison of the values of $\epsilon$ for the folding between experiments and simulations indicates that the indenter friction coefficient would be estimated as $\mu_{\rm i}\simeq 0.2$. We also compare the force-displacement curve of the Unfolding mode between experiments and simulations in Fig.~\ref{fig:2}(c). We show the simulation results for $\mu_{\rm i} = 0.18, 0.20, 0.22$ as the shaded region, which sits close to the experimental ones. This comparison is consistent with our estimate $\mu_{\rm i}\simeq 0.2$. In other words, setting the friction coefficient $\mu_{\rm i} = 0.20$ would be a reasonable choice to reproduce the experimental force displacement curve. Hence, we set the friction coefficients to $\mu_{\rm i} = \mu_{\rm e} = 0.20$ throughout when comparing experimental results with simulations and theory. The experimental force displacement curve for the Unfolding mode drops significantly upon slip, while the force decreases continuously in simulation as found in Fig.~\ref{fig:2}(c). The difference in post-sliding behavior between experiments and simulations stems from differences in contact modeling, which needs to be improved to achieve more predictive modeling.

The deformation modes of the shell are now classified into three distinct states as shown in the phase diagram in Fig.~\ref{fig:3}(a): Folding (blue), Pinning (red), Unfolding (green). We express the experiments with the filled symbols, the simulations with the corresponding open symbols ($(\mu_{\mathrm i},\mu_{\mathrm e}) = (0.2, 0.2)$), and the theoretical analysis with background and lines (See their derivations in Sec.~\ref{sec:Elastica model}). We use the rescaled opening angles as $\tilde{\Phi}\equiv\Phi/\pi$ and $\tilde{\Psi}\equiv\Psi/\pi$ in the following. When $\tilde\Phi$ is large enough, the contact point of the indenter slips in the folding direction (Folding). As $\tilde\Phi$ decreases, the shell does not slip (Pinning). When $\tilde\Psi$ is large, the shell always folds into the hole. Our experiments and simulations are in excellent agreement except near the phase boundaries (derived next). In the next section, we derive the theoretical phase boundaries based on the elastica model.

\section{Elastica model}\label{sec:Elastica model}

We have shown that the deformation modes of the shell are classified into three, depending on the two dimensionless parameters: opening angle of the shell, $\Phi$, and that of the hole, $\Psi$. In this section, we analyze the deformation of the shell using the elastica model, assuming that the deformation is planar and that the contact points (both at the indenter and at the edges of the hole) are immobile. We show that the phase boundaries between three modes (Folding, Unfolding, and Pinning) observed in Fig.~\ref{fig:3}(a) are predictable from our elastica model.

\subsection{Governing equations}

The problem is simplified to that of a naturally-curved two-dimensional beam of radius of curvature, $R$, where the centerline of the shell is parameterized by the arc length $s \in [-L, L]$ as $\bm{r}(s) = \bigl( x(s),\, y(s) \bigr)$. In addition to the Cartesian coordinates, we introduce the bending angle $\theta(s)$, defined as the angle between the tangent to the centerline at arc length $s$ and the $x$-axis (See Fig.~\ref{fig:3}(b) for definitions). From left-right symmetry in the natural configurations, only the right half of the shell, $0 \leq s < L$, is considered. The shell is assumed to be clamped at $s=0$, with its horizontal position fixed, while the vertical displacement at the clamp is free to realize the left-right symmetry of the shell. We choose the $x$ and $y$ axes to be the height of the hole edge and the symmetric axis of the shell, respectively. The shell comes into contact with the edge of the hole at $s = s_c$. The height of the tip measured from the edge of the hole decreases by $\Delta_y$ from $Y (\equiv R(\cos\Psi-\cos\Phi)$) to $Y-\Delta_y$. The tip of the shell $s = L$ touches the indenter, whose boundary condition is assumed to be moment-free for simplicity{~\cite{sano2017slip}}.

We now derive the force and moment balance equation for the beam under the above boundary conditions. 
External forces are applied at the two ends as $\bm{\mathcal{F}}_0 = (\mathcal{F}_{0,x},\, 0)$ at $s = 0$ and $\bm{\mathcal{F}}_L = (\mathcal{F}_{L,x},\, -\mathcal{F}_{L,y})$. 
The contact force at the edge $\bm{\mathcal{F}}_{s_c}$ is determined implicitly from the overall mechanical equilibrium of the shell $\bm{\mathcal{F}}_0 + \bm{\mathcal{F}}_{s_c} + \bm{\mathcal{F}}_L = \bm{0}$. The internal force profile along the centerline, $\bm{F}(s)$, can thus be expressed as
\begin{equation}
    \bm{F}(s) =
    \begin{cases}
        -\bm{\mathcal{F}}_0 & (0 \leq s < s_c),\\
        \bm{\mathcal{F}}_L & (s_c \leq s < L),
    \end{cases}
\end{equation}
from the boundary conditions, where the internal force is piecewise constant~\cite{Audoly2010}. We introduce the moment equilibrium relation from the Kirchhoff equation (along $z$-axis) $\bm{M}'(s) \cdot \bm{\hat{e}}_z + \bigl( \bm{\hat{t}}(s) \times \bm{F}(s) \bigr) \cdot \bm{\hat{e}}_z = 0$ where $\bm{\hat{t}}(s)$ is the unit tangent vector to the centerline, and $\bm{\hat{e}}_z$ are the Cartesian basis vectors in the out-of-plane directions, $\bm{M}(s)=B(\theta'(s)-R^{-1})\bm{\hat e_z}$ is the internal bending moment vector. Substituting the internal force $\bm{F}(s)$ for the Kirchhoff equation at the two sub-intervals $0 \leq s < s_c$ and $s_c \leq s < L$ yields the elastica equation:
\begin{equation}
    B\theta''(s) =
        \begin{cases}
            -\mathcal{F}_{0,x}\,\sin\theta(s) & (0 \leq s < s_c)\\
            \mathcal{F}_{L,x}\,\sin\theta(s) + \mathcal{F}_{L,y}{\cos\theta(s)}. & (s_c \leq s < L)
        \end{cases}
    \label{eq:Elastica equation}
\end{equation}
The governing equation, Eqs.~\eqref{eq:Elastica equation}, is solved under the appropriate boundary, constraint, and continuity conditions. Given that there are two second-order differential equations and three unknown variables, a total of seven conditions are required. The boundary conditions arise from the clamped condition at $(s = 0)$ and the moment-free condition at $(s = L)$, given by
\begin{equation}
    \theta(0) = 0,\quad\theta'(L) = R^{-1}
    \label{eq:bc} 
\end{equation}
respectively. The free-end and contact positions are obtained by integrating the deformed tangent along the inextensible centerline, which yields three constraints:
\begin{align}
    \int_0^{s_c} \hat{\bm{t}} \cdot \hat{\bm{e}}_x \, ds &= R\sin{\Psi}, \\
    \int_{s_c}^L \hat{\bm{t}} \cdot \hat{\bm{e}}_x \, ds &= R(\sin{\Phi}-\sin\Psi), \\
    \int_{s_c}^L \hat{\bm{t}} \cdot \hat{\bm{e}}_y \, ds &= {R(\cos\Psi-\cos\Phi)}-\Delta_y.
    \label{eq:cc}
\end{align}
The solutions for two regions $0<s<s_c$ and $s_c<s<L$ need to be matched at the contact arclength $s = s_c$. In general, when the surface of a slender structure is in frictional contact, one needs to take into account the friction force and torque~\cite{Grandgeorge2021cap}. However, given that our shell is sufficiently slender, we assume that the torque around the centerline due to contact friction is negligible, a reasonable assumption supported by our experimental observations. The continuity conditions require that both the curvature and the deflection angle remain continuous at $ s = s_c $, namely,
\begin{align}
    \lim_{s \to s_c{-0}} \theta(s) &= \lim_{s \to s_c{+0}} \theta(s),\\
    \lim_{s \to s_c{-0}} \theta'(s) &= \lim_{s \to s_c{+0}} \theta'(s).
    \label{eq:continuity_conditions}
\end{align}
The above set of conditions (Eqs.~\eqref{eq:bc}-\eqref{eq:continuity_conditions}) provides seven conditions, which are sufficient to solve our elastica equations (Eqs.~\eqref{eq:Elastica equation}).

We rescale the set of equations by introducing characteristic quantities for length and force. The representative length scale is taken as $R$, and the characteristic force scale is chosen as $B/R^{2}$ (typical force for bending). The rescaled arc length is defined by $\tau = s / R$, where $0 \leq \tau < \Phi (=L/R)$, and the contact position corresponds to $\tau = s_c/R = \Psi$.
The forces appearing in the elastica equation, Eqs.~\eqref{eq:Elastica equation}, are rescaled as follows:
\begin{align}
    f_{0,x} \equiv \frac{\mathcal{F}_{0,x}}{B / R^{2}},\quad 
    f_{L,x} \equiv \frac{\mathcal{F}_{L,x}}{B / R^{2}},\quad
    f_{L,y} \equiv \frac{\mathcal{F}_{L,y}}{B / R^{2}}.
    \label{eq:f nondimensionalize}
\end{align}
By defining the deflection angle in $\tau$ coordinate, $\vartheta(\tau) \equiv \theta(\tau R)$, and denoting differentiation with respect to $\tau$ as $d(\cdot)/d\tau = \dot{(\cdot)}$, we rewrite elastica equation\eqref{eq:Elastica equation} as
\begin{equation}
    \ddot{\vartheta}(\tau) =
    \begin{cases}
    -f_{0,x} \sin \vartheta(\tau) & (0 \leq \tau < \Psi) \\
    f_{L,x} \sin \vartheta(\tau) + f_{L,y} \cos \vartheta(\tau) & (\Psi \leq \tau < \Phi).
    \end{cases}
    \label{eq:Elastic_equation nondimensionalize}
\end{equation}
The boundary conditions are rescaled as
\begin{equation}
    \vartheta(0) = 0,~\dot{\vartheta}(\Phi) = 1.
    \label{eq:bc_res}   
\end{equation}
The constraint conditions are rewritten as
\begin{align}
    \int_0^{\Psi} \cos\vartheta(\tau) d\tau &= \sin{\Psi}, 
    \label{eq:cc1}\\
    \int_{\Psi}^{\Phi} \cos\vartheta(\tau) d\tau &= \sin{\Phi}-\sin\Psi, 
    \label{eq:cc2}\\
    \int_{\Psi}^{\Phi} \sin\vartheta(\tau) d\tau &= {\cos\Psi-\cos\Phi}-\epsilon, 
    \label{eq:cc3}
\end{align}
with the rescaled displacement, $\epsilon = \Delta_y/R$. The continuity conditions are given by
\begin{align}
    \lim_{\tau \to \Psi{-0}} \vartheta(\tau) &= \lim_{\tau \to \Psi{+0}} \vartheta(\tau),\\
    \lim_{\tau \to \Psi{-0}} \dot{\vartheta}(\tau) &= \lim_{\tau \to \Psi{+0}} \dot{\vartheta}(\tau). 
\label{eq:continuity_conditions_res}
\end{align}
The solution of the set of Eqs.~\eqref{eq:Elastic_equation nondimensionalize}-\eqref{eq:continuity_conditions_res} provides the stationary shape of the beam for given system parameters, $(\Phi,\Psi)$, and the rescaled displacement, $\epsilon$.

\subsection{Linear response theory}

The general framework formulated in the previous section yields the exact deformed beam shape. However, according to the experimental observations, the slippage occurs when the displacement is small. Hence, the solution of Eqs.~\eqref{eq:Elastic_equation nondimensionalize}-\eqref{eq:continuity_conditions_res} within the linear response would be sufficient to predict the deformation of the shell. We now solve the elastica equation within the regime of linear response where the rescaled displacement, $\epsilon \equiv \Delta_y / R,$ is sufficiently small as $\epsilon\ll1$. We seek only the leading-order term in the solution of the elastica equation~\cite{yoshida2024indentation}. In the absence of an applied load ($\epsilon = 0 $), the shell is curved with its spontaneous curvature (without any stress), and the deflection angle is given by $\vartheta_0(\tau) = \tau $. When a small load is applied, the deflection angle can be expressed as a perturbation expansion around this unloaded state:
\begin{equation}
    \vartheta(\tau) = \vartheta_0(\tau) + \epsilon\,\vartheta_1(\tau) + O(\epsilon^2).
    \label{eq:theta_expd}
\end{equation}
Within the linear response regime, all forces would be assumed to be proportional to $\epsilon $, i.e.,
\begin{equation}
    f_{0,x} = \bar{k}_x\epsilon,\quad
    f_{L,x} = k_x\epsilon,\quad
    f_{L,y} = k_y\epsilon,
    \label{eq:f_expd}
\end{equation}
where, $\bar{k}_x, k_x$ and $k_y$ represent the stiffness parameter for the corresponding forces. By substituting Eqs.~\eqref{eq:theta_expd} and \eqref{eq:f_expd} into the rescaled governing equations Eqs.~\eqref{eq:Elastic_equation nondimensionalize}-\eqref{eq:continuity_conditions_res} and leaving the $O(\epsilon)$ terms only (See Appendix for details), we arrive at the matrix equations for stiffness parameters, $(\bar{k}_x, k_x, k_y)$, as:
\begin{equation}
    \left( \begin{array}{c} \bar{k}_x\\ k_x \\ k_y \end{array} \right) = \mathbf{C}^{-1}~\left( \begin{array}{c} 0\\ 0 \\ -1 \end{array} \right),
    \label{eq:k_linear}
\end{equation}
where 3$\times$3 matrix, $\mathbf{C} = \mathbf{C}(\Psi,\Phi)$, are solely functions of $(\Psi, \Phi)$ (See their definitions in Appendix). In other words, given a set of system parameters, $(\Psi, \Phi)$, we can compute the stiffness parameters $(\bar{k}_x, k_x, k_y)$ from Eqs.~\eqref{eq:k_linear}.

\subsection{Classifications of deformation modes}

The set of coefficients ($\bar{k}_x,k_x,k_y$) derived in the previous section characterizes the slip condition of the shell. When the free-end slips, the contact forces at $s = L$, $(\mathcal{F}_{L,x}, \mathcal{F}_{L,y})$ would satisfy the Amontons-Coulomb friction law as
\begin{equation}
    \pm \mu_{\mathrm{i}}=\frac{\mathcal{F}_{L,x}}{\mathcal{F}_{L,y}}=\frac{k_{x}}{k_{y}}  .
    \label{eq:mu_i theory}
\end{equation}
where the sign of the coefficient determine whether the shell ``closes (Folding)'' or ``opens (Unfolding)''. On the other hand, the Amontons-Coulomb law is satisfied when the shell slips at the contact point with the edge. The contact force $\bm{\mathcal{F}}_{s_c}$ would be obtained from mechanical equilibrium as $\bm{\mathcal{F}}_{s_c}=-\bm{\mathcal{F}}_{L}-\bm{\mathcal{F}}_{0}$. The tangential and normal components, $\mathcal{F}_{s_c,t}$ and $\mathcal{F}_{s_c,n}$ will be obtained as $\mathcal{F}_{s_c,t}=\bm{\mathcal{F}}_{s_c}\cdot\bm{\hat{t}}(s_c)$ and $\mathcal{F}_{s_c,n}=\bm{\mathcal{F}}_{s_c}\cdot\bm{\hat{n}}(s_c)$, respectively. In linear response theory, the deformed tangent, $\bm{\hat{t}}(s_c)$, and normal, $\bm{\hat{n}}(s_c)$, here correspond to those in the reference configuration. Thus, the slip condition on the edge becomes
\begin{equation}
    \mu_{\mathrm{e}}=\frac{\mathcal{F}_{s_c,t}}{\mathcal{F}_{s_c,n}}=\frac{-(k_x+\bar{k}_x)\cos\Psi+k_y\sin\Psi}{(k_x+\bar{k}_x)\sin\Psi+k_y\cos\Psi}.
    \label{eq:mu_e theory}
\end{equation}
Note that the shell slips on the edge only in the opposite direction along the tangent, because $\mathcal{F}_{s_c,t} > 0$ holds within our experimental parameter space $(\tilde{\Psi}, \tilde{\Phi})$. 
Given the values of the friction coefficients $\mu_{\mathrm{i}}$ and $\mu_{\mathrm{e}}$, Eqs.~\eqref{eq:mu_i theory} and \eqref{eq:mu_e theory} provide curves on $\Psi - \Phi$ plane, which correspond to the analytical representations of the phase boundaries in Fig.~\ref{fig:3}(a). We plot our analytical predictions based on Eqs.~\eqref{eq:mu_i theory} and \eqref{eq:mu_e theory} when the friction coefficient $\mu_{\mathrm{i}}=\mu_{\mathrm{e}} = 0.2$, which are consistent with both experiments and simulations (Fig.~\ref{fig:3}(a)). In summary, we are successful in classifying the sliding modes of the shell when compressed between a flat indenter and a hole, based on our elastica model within the linear response $\epsilon\ll1$.

\section{Effects of friction}\label{sec:Effects of friction}

\begin{figure}[!b]
    \centering
    \includegraphics[width=\linewidth]{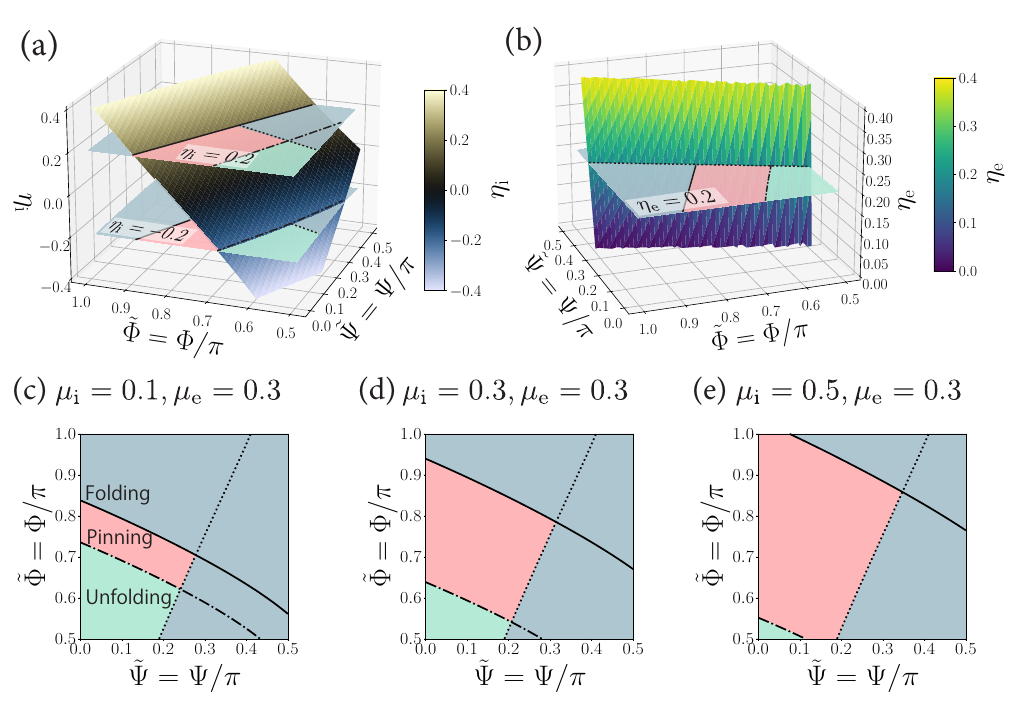}
    \caption{
        The effect of friction coefficients on the sliding behavior of the strip based on linear response theory. The force ratio at (a) $s = L$ and (b) $s = s_c$ predicted from the linear response as a function of $(\tilde\Psi,\tilde\Phi)$. (a) The intersections of the surface with the planes $\eta_{\mathrm i}=\pm\mu_{\mathrm i}=\pm0.2$ correspond to the slip conditions at the indenter (either opening or closing the tip, respectively). (b) The intersection with the plane $\eta_{\mathrm e}=\mu_{\mathrm e}=0.2$ describes the slip condition at the edge of the hole. (c)–(e) Phase diagrams constructed from linear response theory for different friction coefficients: (c) $\mu_{\mathrm e}=0.3,~\mu_{\mathrm i} = 0.1$, (d) $\mu_{\mathrm e}=0.3,~\mu_{\mathrm i}= 0.3$, and (e) $\mu_{\mathrm e}=0.3,~\mu_{\mathrm i}= 0.5$.
    }
    \label{fig:4}
\end{figure}

We have classified the deformation modes of shells based on the elastica model as Eqs.~\eqref{eq:mu_i theory} and \eqref{eq:mu_e theory}, where we have fixed the friction coefficients $\mu_{\mathrm i}$ and $\mu_{\mathrm e}$ throughout. In this section, we discuss the effects of friction coefficients by applying the analytical formulas. The right-hand sides of Eqs.~\eqref{eq:mu_i theory} and \eqref{eq:mu_e theory} provide the force ratios of the contact points as $\eta_{\mathrm i}(\tilde\Psi,\tilde\Phi) \equiv \mathcal{F}_{L,x}/\mathcal{F}_{L,y}$ and $\eta_{\mathrm e}(\tilde\Psi,\tilde\Phi) \equiv \mathcal{F}_{s_c,t}/\mathcal{F}_{s_c,n}$, respectively for a given set of rescaled opening angles $(\tilde\Psi,\tilde\Phi)$. We show in Fig.~\ref{fig:4}(a) and (b) the surfaces of the force ratios, $\eta_{\mathrm i}$ and $\eta_{\mathrm e}$ that are uniquely determined for given dimensional parameters $\Phi$ and $\Psi$. 
The force ratios are calculated for each parameter $(\tilde\Psi,\tilde\Phi)$ and plotted as a point cloud in 3D coordinates $(\tilde\Psi,\tilde\Phi,\eta_\alpha)$, which are smoothly interpolated to construct a continuous surface.

The contour lines of the surfaces in Fig.~\ref{fig:4} (a) and (b) give the phase boundaries in the phase diagram for given friction coefficients $\mu_{\mathrm i}$ and $\mu_{\mathrm e}$. In other words, we can draw the phase diagrams for different coefficients by considering the intersections as $\pm\mu_{\mathrm i} = \eta_{\mathrm i}$ and $\mu_{\mathrm e}= \ \eta_{\mathrm e}$. Here, $+\mu_{\mathrm i} = \eta_{\mathrm i}$ indicates slip in the negative direction of the force, and $-\mu_{\mathrm i} = \eta_{\mathrm i}$ indicates slip in the positive direction of the force on the indenter. The phase boundaries in Fig.~\ref{fig:3}(a) could be reproduced as follows. The planes drawn in Fig.~\ref{fig:4}(a) are the planes for $\eta_{\mathrm i}= \mp 0.2$, which correspond to the phase boundaries for sliding in the $x$-direction (Unfolding-Pinning boundary) and the opposite $x$-direction (Folding-Pinning boundary), respectively. On the other hand, the plane drawn in Fig.~\ref{fig:4}(b) is the plane for $\eta_{\mathrm e} = 0.2$, which represents the threshold for sliding at the edge in the tangent direction (Folding-Unfolding and Folding-Pinning boundary). We construct the phase diagram for different friction coefficients based on the above strategies. The surface plots of the force ratios $\eta_{\mathrm i}$ and $\eta_{\mathrm e}$ indicate that the change of $\mu_{\mathrm i}$ would affect the phase diagram significantly. In contrast, the phase boundary due to the slip at the edge is insensitive to $\mu_{\mathrm e}$ because the contour line of $\mu_{\mathrm e}= \eta_{\mathrm e}$ does not change essentially. We plot the phase diagrams of several sets of friction coefficients $(\mu_{\mathrm i}, \mu_{\mathrm e})$ as $(\mu_{\mathrm i}, \mu_{\mathrm e}) = (0.1,0.3), (0.3, 0.3), (0.5,0.3),$ respectively in FigS.~\ref{fig:4}(c)-(e), finding that the pinning mode is enhanced for larger friction coefficients at the indenter, which is consistent with our intuition.

\section{Discussion and Summary}\label{sec:Discussion and Summary}

We investigated the deformation of an open cylindrical shell during insertion into a hole by combining experiments, simulation, and theory. 
We systematically classified the deformation modes, using the geometric parameters $(\Psi,\Phi)$ and friction coefficients~$(\mu_{\mathrm i},\mu_{\mathrm e})$. Three characteristic modes (Folding, Pinning, and Unfolding) were identified, and their phase boundaries were quantitatively determined. Experiments, numerical simulations, and theoretical analysis based on the elastica model are in agreement with each other~\cite{alldata}.

We have carefully conducted experiments to ensure that asymmetric sliding and rolling are not observed. Nevertheless, asymmetric deformation of the tips or out-of-plane deformations of the shell could happen if misalignment of the shell or that of the contact condition is significant. The critical condition at which asymmetric deformations emerge could be investigated by introducing asymmetry or defects into the simulation, which we leave for future study. The prediction of asymmetric sliding mode would be relevant when we apply our study to many body systems, such as a randomly stacked system~\cite{sano2023randomly}. 

Lastly, we note that friction at the indenter and at the edge contribute differently to sliding behavior. In Fig.~\ref{fig:4}, we have analyzed the phase boundaries for different friction coefficients by identifying the intersections of the surfaces. The value of $\mu_{\mathrm e}$ does not affect the locations of the phase boundaries for the slip at the edge on the $(\Phi,\Psi)$ plane significantly, while that of $\mu_{\mathrm i}$ affects a lot, as can be seen from the shape of the surfaces in Fig,~\ref{fig:4}(a)(b). In other words, the precision of the indenter friction coefficient, $\mu_{\mathrm i}$, is more critical in predicting the sliding behavior than the edge friction coefficient, $\mu_{\mathrm e}$. The difference in friction coefficients at the tips of the shell could contribute more to the asymmetric sliding or rolling than at the edge contacts.

\section*{Acknowledgement}

We acknowledge the funding from MEXT KAKENHI 24H00299 (T.G.S.), JST FOREST Program, Grant Number JPMJFR212W (T.G.S.), and the 2025 research grant from the Amano Institute of Technology (K. Y.).

\appendix

\section{Derivation of Linear Response}
We provide the derivation of the linear response theory under the small parameter $\epsilon=\Delta_y/R\ll 1$, where we arrive at the results of the linear response theory, Eq.~\eqref{eq:k_linear}, from Eq.~\eqref{eq:Elastic_equation nondimensionalize}. We first express the bending angles  deviated from the natural configuration, $\vartheta_1(\tau)$ (See Eq.~\eqref{eq:theta_expd}) before and after the contact at $\tau = \Psi$ as:
\begin{equation}\vartheta_1(\tau)= 
    \begin{cases}
        \vartheta_1 ^-(\tau)&(0\leq\tau<\Psi)\\ 
        \vartheta_1 ^+(\tau)&(\Psi\leq\tau<\Phi).
    \end{cases}
\end{equation}
Below, we construct the first-order solution for each region, $\vartheta_1 ^{\pm}(\tau)$, by applying a perturbative approach to Eq.~\eqref{eq:Elastic_equation nondimensionalize}.

\subsection{First-order solution in $0\leq\tau<\Psi$}

By substituting $\vartheta(\tau) = \tau + \epsilon\vartheta_1 ^-(\tau)$ and $f_{0,x} = \bar{k}_x\epsilon$ into Eq.~\eqref{eq:Elastic_equation nondimensionalize}, we rewrite Eq.~\eqref{eq:Elastic_equation nondimensionalize} in the region for $0\leq\tau<\Psi$ as:
\begin{equation}
    \ddot{\vartheta}^{-}_1(\tau) = -\bar k_x\sin\tau
    \label{eq:ddtheta1m}
\end{equation}
with the boundary conditions $\vartheta_1(0)=0$, $\dot\vartheta_{1}^{-}(\Psi)=\dot\vartheta_{\mathrm e}$, where $\dot\vartheta_{\mathrm e}$ represents the curvature at the edge. Note that the curvature at the edge $\dot{\vartheta}_{\mathrm e}$ would be determined later from the matching condition with the solution of $\vartheta_1^+$. The solution satisfying the boundary conditions is obtained by integrating Eq.~\eqref{eq:ddtheta1m} twice as
\begin{equation}
    \vartheta_{1}^{-}(\tau) = -\bar k_x\{\tau\cos\Psi-\sin\tau\}+\dot\vartheta_{\mathrm e}\tau.
    \label{eq:theta^-_1}
\end{equation}

\subsection{First-order solution in $\Psi\leq\tau<\Phi$}

The governing equation in the region for $\Psi\leq\tau<\Phi$ is calculated by substituting $\vartheta(\tau) = \tau + \epsilon\vartheta_1 ^+(\tau)$, $f_{L,x} = {k}_x\epsilon$, and $f_{L,y} = k_y\epsilon$ into Eq.~\eqref{eq:Elastic_equation nondimensionalize} as
\begin{equation}
    \ddot{\vartheta}^{+}_1(\tau) = k_x\sin\tau + k_y\cos\tau 
    \label{eq:ddthetap}
\end{equation}
with the boundary conditions $\vartheta^{+}_1(\Psi)=\vartheta_{\mathrm e}$, $\dot\vartheta^{+}_1(\Phi)=0$. Integrating Eq.~\eqref{eq:ddthetap} once within the interval $\tau<\tau'<\Phi$  and imposing one of the boundary conditions, $\dot\vartheta^{+}_1(\Phi)=0$, yields
\begin{equation}
    \dot\vartheta_1^{+}(\tau)=k_x(\cos\Phi-\cos\tau)-k_y(\sin\Phi-\sin\tau),
    \label{eq:dottheta^+_1}
\end{equation}
which allows us to obtain the edge curvature from the continuity condition at $\tau = \Psi$ (Eq.~\eqref{eq:continuity_conditions_res}) as $\dot\vartheta_{\mathrm e}  = \lim_{\tau\to\Psi+0}\dot\vartheta_1^{+}(\tau) = k_x(\cos\Phi-\cos\Psi)-k_y(\sin\Phi-\sin\Psi)$. The bending angle $\vartheta_1^{+}(\tau)$ is determined by integrating the curvature $\dot\vartheta_1^{+}(\tau)$ with the remaining boundary condition,$\vartheta_1 ^+(\Psi) = \vartheta_{\mathrm e}$. 
Therefore, we get 
\begin{equation}
    \begin{aligned}
    \vartheta_1^{+}(\tau) = 
     &k_x [\tau\cos \Phi - \sin \tau - (\Psi\cos \Psi - \sin \Psi) ] \\
    - &k_y [ \tau\sin \Phi + \cos \tau - (\Psi\sin \Psi + \cos \Psi )] \\
    - &\bar{k}_x [ \Psi\cos \Psi - \sin \Psi ]
    .
    \end{aligned}
   \label{eq:theta^+_1}
\end{equation}
We have successfully obtained the 1st order solution $\vartheta_1 ^{\pm}(\tau)$ as Eqs.~\eqref{eq:theta^-_1} and \eqref{eq:theta^+_1}. Our remaining task is to determine the stiffness parameters, $k_x$, $k_y$, and $\bar{k}_x$ by employing the inextensible constraints Eqs.~\eqref{eq:cc1}-\eqref{eq:cc3}.

\subsection{Derivation of $k_x,k_y,\bar{k}_x$}

Our goal is to formulate a system of equations to find $k_x, k_y, \bar{k}_x$ using the geometric parameters, $\Psi$ and $\Phi$. The trigonometric functions used in the inextensible constraints, Eqs.~\eqref{eq:cc1}-\eqref{eq:cc3}, can be rewritten as follows by extracting their $O(\epsilon)$ contributions:
\begin{align}
    \sin(\vartheta_0 + \epsilon \vartheta_1) &= \sin\vartheta_0 + \epsilon\vartheta_1\cos\vartheta_0 + O(\epsilon^2)
    \label{eq:trigonometric function approximation sin},\\
    \cos(\vartheta_0 + \epsilon \vartheta_1) &= \cos\vartheta_0 - \epsilon\vartheta_1\sin\vartheta_0 + O(\epsilon^2).
    \label{eq:trigonometric function approximation cos}
\end{align}
The constraints in Eqs.~\eqref{eq:cc1}, \eqref{eq:cc2}, and \eqref{eq:cc3} can be rewritten as the following integral conditions by substituting Eqs.~\eqref{eq:trigonometric function approximation sin}, \eqref{eq:trigonometric function approximation cos} and collecting $O(\epsilon)$ only:
\begin{align}
    \int_0^\Psi \vartheta_1^{-}(\tau)\sin\tau\,d\tau&= 0,
    \label{eq:cc_liner_1}\\
    \int_\Psi^\Phi \vartheta_1^{+}(\tau)\sin\tau\,d\tau&= 0, 
    \label{eq:cc_liner_2}\\
    \int_\Psi^\Phi \vartheta_1^{+}(\tau)\cos\tau\,d\tau&= -1.
    \label{eq:cc_liner_3}
\end{align}
We substitute Eqs.~\eqref{eq:theta^-_1} and \eqref{eq:theta^+_1} into Eqs.~\eqref{eq:cc_liner_1}-\eqref{eq:cc_liner_3}. Notice that the coefficients, $k_x,k_y$ and $\bar{k}_x$ are linear in Eqs.~\eqref{eq:theta^-_1} and \eqref{eq:theta^+_1} ($\vartheta_1 ^{\pm}$) is also linear in Eqs.~\eqref{eq:cc_liner_1}-\eqref{eq:cc_liner_3}. Hence, we get the set of three linear equations as
\begin{equation}
    \mathbf{C}\left( \begin{array}{c} \bar{k}_x\\ k_x \\ k_y \end{array} \right) = \left( \begin{array}{c} 0\\ 0 \\ -1 \end{array} \right).
    \label{eq:App liner system}
\end{equation}
where $\mathbf{C}=\mathbf{C}(\Psi,\Phi)$ is a $3\times 3$ matrix. Equation \eqref{eq:App liner system} is readily rewritten as Eq.~\eqref{eq:k_linear} in the main text.

The explicit expressions for the matrix elements $C_{\alpha\beta}$ appearing in Eq.~\eqref{eq:App liner system} are summarized below:
\begin{align}
    C_{11}&=\int_0^\Psi (\tau\cos\Psi-\sin\tau)\sin\tau d\tau,\\
    C_{12} &= \int_0^\Psi \tau(\cos\Phi-\cos\Psi)\sin\tau d\tau,\\
    C_{13} &= \int_0^\Psi \tau(\sin\Phi-\sin\Psi)\sin\tau d\tau,\\
    C_{21} &= \int_{\Psi}^{\Phi} (\Psi\cos\Psi-\sin\Psi)\sin \tau\,d\tau, \\
    C_{22} &= \int_{\Psi}^{\Phi} \left[ \Psi(\cos\Phi-\cos\Psi) +\tau\cos\Phi-\sin\tau\right]\sin \tau\,d\tau, \\
    C_{23} &= \int_{\Psi}^{\Phi} \left[\Psi(\sin\Phi-\sin\Psi)+\tau\sin\Phi+\cos\tau\right]\sin \tau\,d\tau, \\
    C_{31} &= \int_{\Psi}^{\Phi} (\Psi\cos\Psi-\sin\Psi)\cos \tau\,d\tau, \\
    C_{32} &= \int_{\Psi}^{\Phi}\left[ \Psi(\cos\Phi-\cos\Psi) +\tau\cos\Phi-\sin\tau\right]\cos \tau\,d\tau, \\
    C_{33} &= \int_{\Psi}^{\Phi}\left[\Psi(\sin\Phi-\sin\Psi)+\tau\sin\Phi+\cos\tau\right]\cos \tau\,d\tau.
\end{align}
\bibliography{apssamp}
\end{document}